\documentclass[twocolumn,showpacs,showkeys,preprintnumbers,amsmath,amssymb]{revtex4}
\usepackage{graphicx}% Include figure files
\usepackage{dcolumn}% Align table columns on decimal point
\usepackage{bm}% bold math
\usepackage{epsfig}
\begin{document}

\title{Antipersistant Effects in the Dynamics of a Competing Population}% Force line breaks with \\

\author{K. H. Lee and K. Y. Michael Wong}
\email{eyeshadow.lee@gmail.com, phkywong@ust.hk}

\affiliation{
Department of Physics, Hong Kong University of Science and Technology, Hong Kong, China
}

\date[]{Received Date Month Year}

\begin{abstract}

We consider a population of agents competing for finite resources
using strategies based on two channels of signals. The model is
applicable to financial markets, ecosystems and computer networks.
We find that the dynamics of the system is determined by the
correlation between the two channels. In particular, occasional
mismatches of the signals induce a series of transitions among
numerous attractors. Surprisingly, in contrast to the effects of
noises on dynamical systems normally resulting in a large number of
attractors, the number of attractors due to the mismatched signals
remains finite. Both simulations and analyses show that this can be
explained by the antipersistent nature of the dynamics.
Antipersistence refers to the response of the system to a given
signal being opposite to that of the signal's previous occurrence,
and is a consequence of the competition of the agents to make
minority decisions. Thus, it is essential for stabilizing the
dynamical systems.

\end{abstract}



\maketitle

\section{\label{sec:level1} Introduction}

Starting from the last decade, physicists have been trying to cope
with the issues traditionally approached by economists using their
own tools and methodologies~\cite{stanley}. This research has been
dubbed `econophysics'. Since the financial market can be considered
as a complex system with large population, techniques of statistical
mechanics are applicable. The Minority Game (MG) is a simple
statistical mechanical model for us to study complex economical
systems. The minority-winning nature of MG is close to situations in
daily life. For example, in financial markets, all buyers and
sellers want to maximize their wealth. Only the minority side gains
the benefits because of excess demand or supply of the whole system.
This often creates the {\it antipersistent} behavior, which refers
to agents rushing to the minority side as adaptive attempts in the
evolving environment, resulting in opposite responses to the
consecutive occurrence of a given signal~\cite{Marsili}.
%%Similar
%%behavior can be found in the stock market.

Decisions of agents in the market are often responses to the
information relevant to them. However, noisy information may cause
confusion in their decision making. Hence, we are interested in
studying the effects of noisy signals to the system and the
underlying physics. In order to model these features, a new MG
model, ``The Minority Game with Errors'' with two channels of
signals is introduced in this paper.

%%In Section~\ref{sec:level2}, the new model of MG will be introduced
%%in detail. Computer simulation results of this model will be shown
%%in Section~\ref{sec:level3}. In
%%Section~\ref{sec:level4}~and~\ref{sec:level5}, we will analyze this
%%complex system using different tools and approaches. Finally, a
%%conclusion will be given in the Section~\ref{sec:level6}.

%%%%%%%%%%%%%%%%%%%%%%%%%%%%%%%%%%%%%%%%%%%%%%%%%%%%%%%%%%%%%%%%%%%%%%

\section{\label{sec:level2} The Minority Game with Errors}

This model is based on the original Minority Game~\cite{Challet}.
``The Minority Game with Errors'' is a binary game that $N$ ($N$
must be odd) agents have to make binary decisions (denoted by 0 or
1) independently at every time step and the agents making the
minority decision win. The decisions of the agents are based on {\it
signals}. In the original version of MG, the signals are {\it
endogenous}, which consist of the winning bits of the most recent
$M$ steps. In another version of the game, the signals are {\it
exogenous}~\cite{Cavagna}, which are randomly selected from $D$
signals at each step. Studies had shown that the behavior of these
two versions are very similar~\cite{Cavagna}. In this paper, we
adopt the exogenous version of MG.

\begin{figure}
\centerline{\epsfig{figure=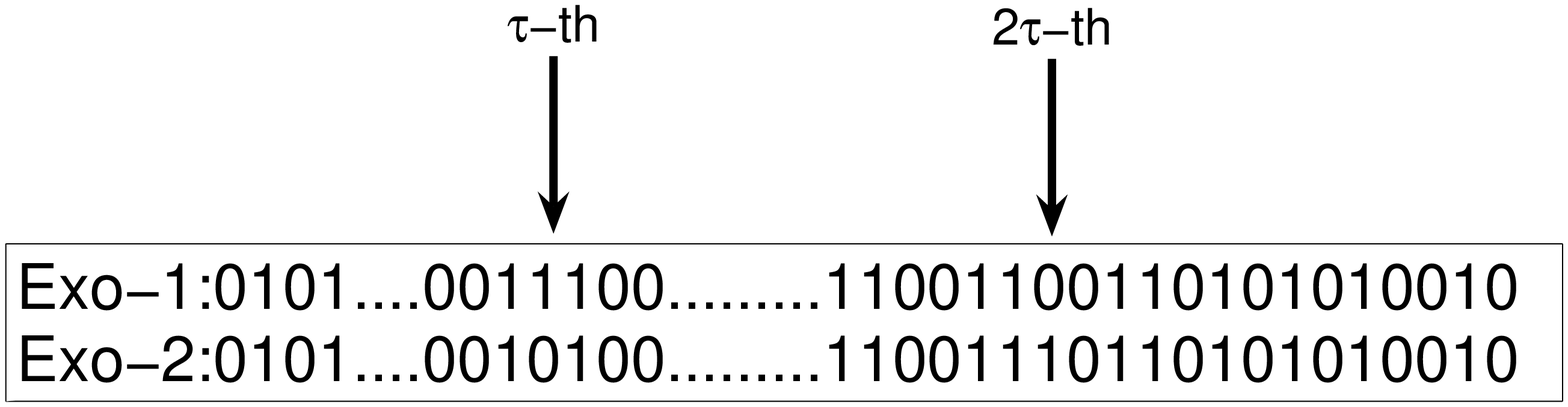,width=1.0\linewidth}}
\caption{\label{twoCH} Illustration of the signal for $D=2$, with
errors for every $\tau$ time steps. Exo-1 is randomly generated.
Exo-2 is {\it nearly identical} to Exo-1.}
\end{figure}

Each agent holds $s$ strategies which are binary functions mapping
the $D$ signals to decision 0 or 1.They adopt the most successful
one among the $s$ strategies in order to have a higher chance to
win. The success of a strategy is measured by its cumulative payoff
(or virtual point), which changes by +1 for a minority decision and
-1 for a majority decision. In order to model the errors in signals,
we introduce $s$ channels of external {\it nearly identical}
signals, each feeding the corresponding strategy. For $s=2$, we call
the signals Exo-1 and Exo-2. In detail, Exo-1 is a randomly
generated signal. Exo-2 is the same as Exo-1 in most of the time
steps, but differs occasionally, as shown in Fig.~\ref{twoCH}. Thus,
we can treat them as one channel of signals with occasional errors.

%%%%%%%%%% Figure 1 %%%%%%%%%%%%
%%%%%%%%%%%%%%%%%%%%%%%%%%%%%%%%

\section{\label{sec:level3} Numerical Results}

\begin{figure}
\centerline{\epsfig{figure=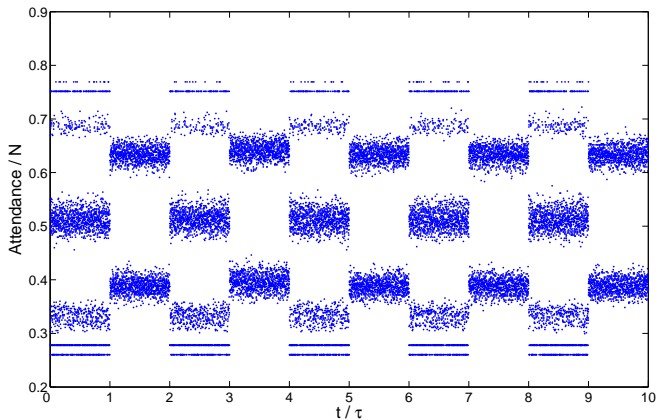,width=1.0\linewidth}}
\caption{\label{sudden} Time series of attendance in a simulation
with $N=511$, $D=2$, $s=2$ and $\tau=10^4$.}
\end{figure}

We schedule the errors to occur once every $\tau$ steps, where
$\tau$ is a large number compared with the period of dynamics so
that agents can have enough time to adapt to changes of the system
caused by errors. Figure~\ref{sudden} shows a simulated time series
of {\it attendance}, which is defined as the number of agents
choosing 1 in a time step. We can observe that attendance clusters
only on several values (around 0.3, 0.5 and 0.7) before the
occurrence of the first error at $t=\tau$. Using the language of
non-linear science, there exists an attractor which contributes to
the dynamics of the system. After an error occurs, the attendance
clusters on another set of values (around 0.4 and 0.6). Hence, a new
attractor is formed due to the errors in the input signals. After
each occurrence of errors, the pattern of attendance alternates
between the two attractor states of the system. Other intervals between errors, as well as other samples, result in
two further attractors. Since the error generation mechanism is
random, it is surprising that errors result in regular patterns. In
contrast, many complex systems, when fed with occasional noisy
signals, move among numerous attractors, and therefore cannot
maintain their stability.

From the above results, this system appears to be resistive to huge
changes. Then, we may ask the following questions in order to know
more about its stability:

\begin{enumerate}
\item What is the number of attractors in this system?
\item If the number of attractors is finite, what is the mechanism forbidding the occurrence of numerous attractors?
\end{enumerate}

Detailed analyses in the following sections are essential to
answering the above questions.

%%%%%%%%% Figure 2 %%%%%%%%%%%%
%%%%%%%%%%%%%%%%%%%%%%%%%%%%%%%%

\section{\label{sec:level4} Phase Space Analysis}

In order to understand the dynamics of the system easily, $A^{\mu}$
is defined as follows~\cite{Michael,Long}.

\begin{eqnarray}
A^{\mu}(t)=\frac{N_1^{\mu}(t)-N_0^{\mu}(t)}{N},
\end{eqnarray}
where $N_1^{\mu}(t)$, $N_0^{\mu}(t)$ are, respectively, the number
of agents choosing 1 and 0 when the input signal is $\mu$ at time
$t$. For simplicity, we consider $D=2$ so that $\mu$ = 0 or 1. If
$A^{\mu}$ is positive, the minority side will be 0, and vice versa.
Thus, $A^{\mu}$ is a useful tool to analyze the dynamics inside the
system when it responds to different input signals $\mu$. As there
are only $A^0$ and $A^1$, using the two-dimensional phase space is
the most suitable way to describe the dynamics~\cite{Michael}.

%%%%%%%%% Figure 3a,b %%%%%%%%%%%%
\begin{figure}
\centerline{\epsfig{figure=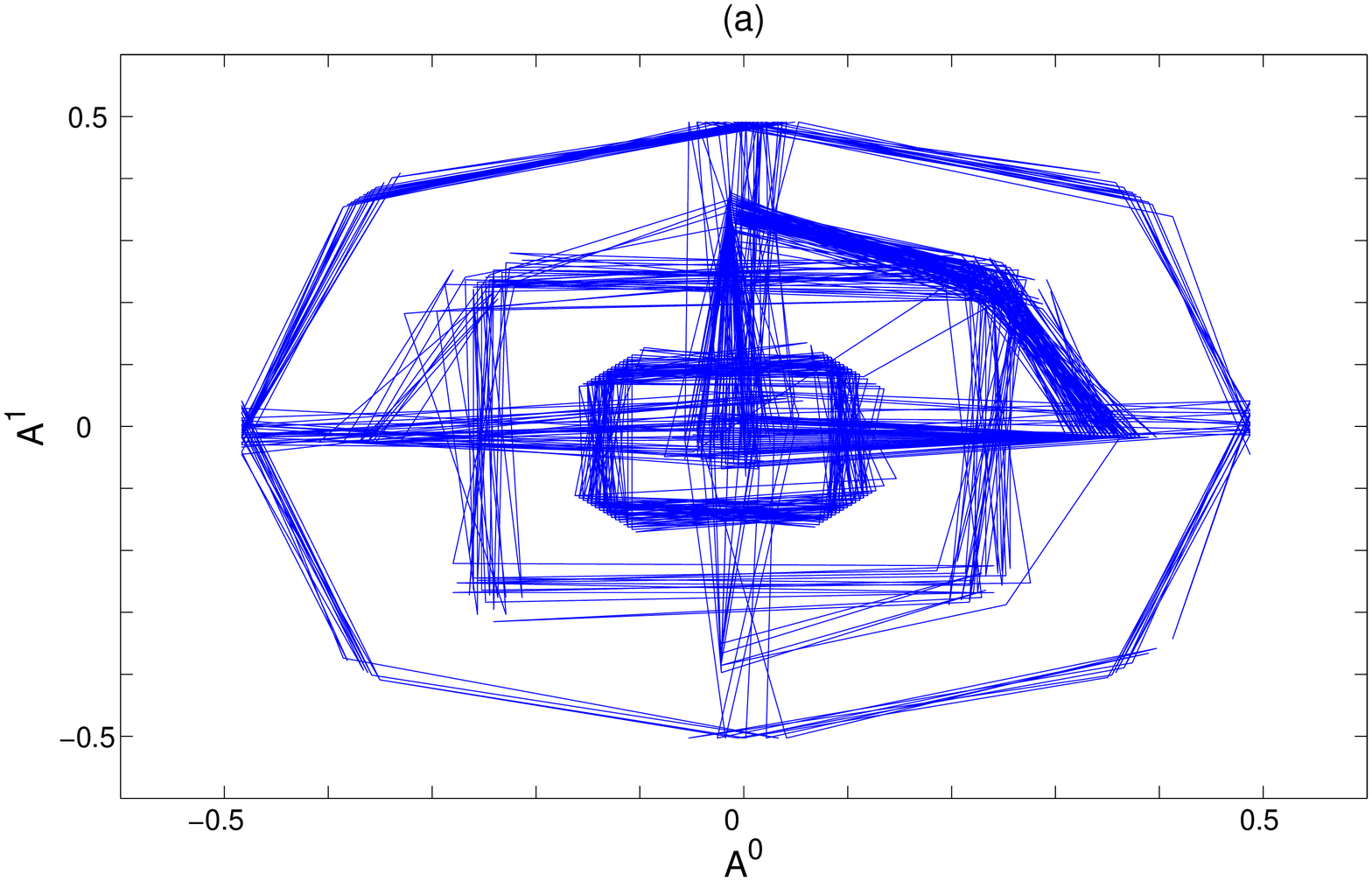,width=1.0\linewidth}}
\centerline{\epsfig{figure=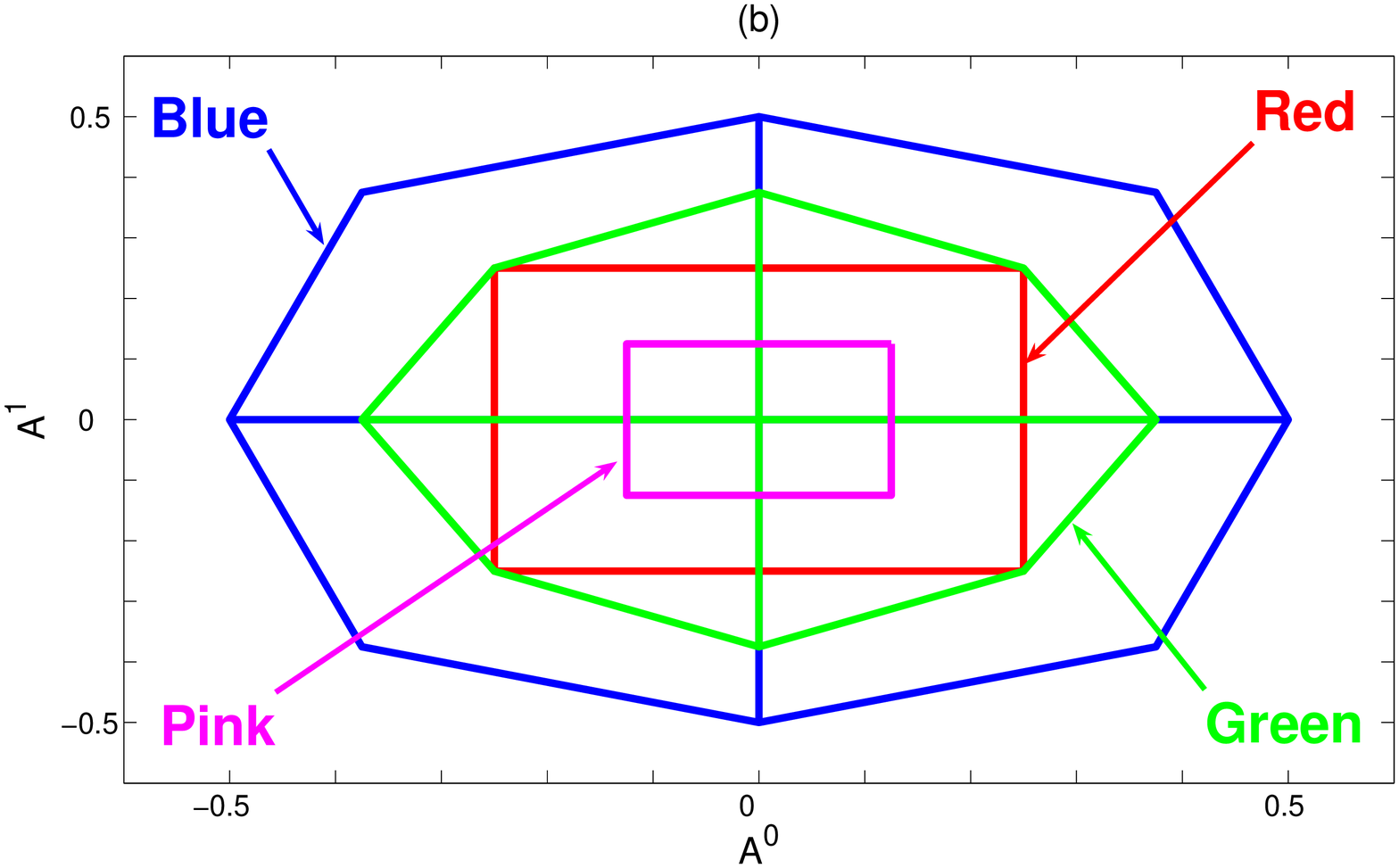,width=1.0\linewidth}}
\caption{\label{phase} (a) Numerical results of the attractors in
the phase space. Parameters: $N=511$, $D=2$, $s=2$. (b) Analytic
calculation in the phase space.}
\end{figure}
%%%%%%%%%%%%%%%%%%%%%%%%%%%%%%%%

From the simulation results in Fig.~\ref{phase}(a), the clustered
data points show that the system stays in different attractors
between the occurrence of error signals. Each attractor has a
polygonal pattern of its own size in the phase space.

Since the dimension is not high ($D=2$), the dynamics of the system
can be studied analytically. For each strategy combination, the
number of agents holding it is assumed to be the same if $N$ is
large compared with the total number of strategy combinations.
Therefore, the attractor states can be calculated by considering
cumulative payoffs of all strategies. The values of these attractor
states in different stages are listed in Table~\ref{attstate}.

%%%%% Table 1 %%%%%
\begin{table}[htp]
\begin{center}

\begin{tabular}{|c|c|c|c|}
  \hline
  Sequence & Colour & Coordinates of attractors & No. of states \\

   &  & ($A^0$,$A^1$) &  \\\hline
  1 & Blue & ($0^{\pm},0^{\pm}$) ($0,\pm1/2$),  & 9 \\
    &      & ($\pm1/2,0$), ($\pm3/8$, $\pm3/8$) & \\\hline
  2 & Red & ($\pm1/4$, $\pm1/4$) & 4 \\\hline
  3 & Green & ($0^{\pm},0^{\pm}$), ($0,\pm3/8$),  & 9 \\
    &       & ($\pm3/8,0$), ($\pm1/4$, $\pm1/4$) & \\\hline
  4 & Pink & ($\pm1/8$, $\pm1/8$) & 4 \\\hline
\end{tabular}

\end{center}
\caption{\label{attstate} Analytical results for different
attractors}
\end{table}
%%%%%%%%%%%%%%%%%%%

In Fig.~\ref{phase}(b), the colour lines are the analytical results
in Table~\ref{attstate}. Before the first error, the attractor of
the system is the blue one. When an error occurs, the attractor
transits to the red one with a smaller size. When a further error
appears, the system restores back to the original blue attractor
with a very high probability, and transits to the green one with a
low probability. Similarly, transitions to further attractors may
continue but with a lower and lower probability. The transitions of
the attractors follow by the sequence as shown in
Table~\ref{attstate}. The pink attractor, the smallest among all,
represents the final stage of attractor transitions. An occurrence
of error in the pink attractor cannot cause it to transit further
into another attractor. It must restore back to the previous
attractor (green) when one more error occurs.

\section{\label{sec:level5} payoff space analysis}

In order to study the mechanism of this stoppage of further
attractor transitions, one should study the {\it payoff space},
whose dimensions are $k_{\mu}(t)$, defined as the number of wins
minus losses of decision 1 up to time $t$ when the game responded to
signal $\mu$. Since a location in the payoff space records the
winning history of decision 1, the cumulative payoffs of all
strategies can be calculated~\cite{Michael}. For example, the
cumulative payoff $\Omega_a(t)$ of strategy $a$ up to time $t$ can
be computed as follow,

\begin{equation}
        \Omega_a(t)=\sum_\mu k_\mu(t)(2\sigma_a^\mu-1),
\label{payoff}
\end{equation}
where $\sigma_a^\mu$ is the binary decision (0 or 1) of strategy $a$
for signal $\mu$. Thus, the dynamics of the system can be totally determined after
knowing the trajectories in the payoff space.

%%%%%%%%% Figure 4a-d %%%%%%%%%%%%
\begin{figure}
\centerline{\epsfig{figure=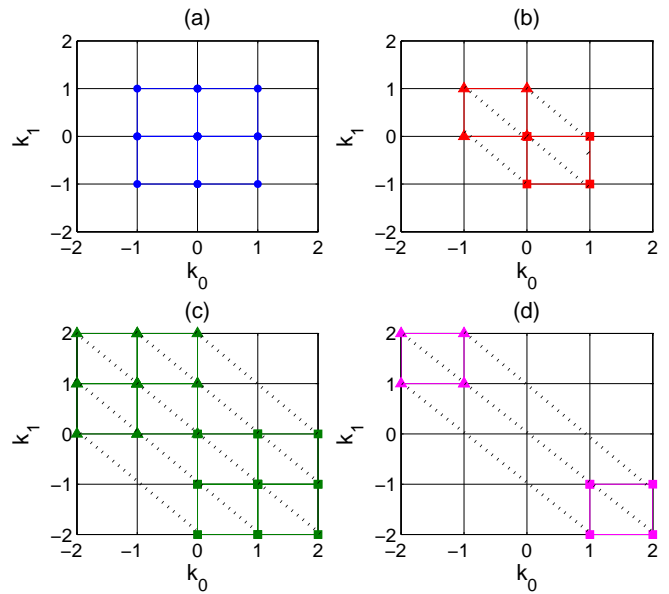,width=1.0\linewidth}}
\caption{\label{4katt} (a) The first attractor (blue), (b) The
second attractor (red), (c) The third attractor (green), (d) The
final attractor (pink).}
\end{figure}
%%%%%%%%%%%%%%%%%%%%%%%%%%%%%%%%

Attractor transitions in the payoff space shown in
Figs.~\ref{4katt}(a)-(d) are the same as those in the phase space
using identical colours and sequences. In Fig.~\ref{4katt}(a), the
nine points represent the nine states of the blue attractor. When an
error occurs in the blue attractor, each state splits into two, due
to different signals in Exo-1 and Exo-2. In order to identify the
two groups of attractor states in Fig.~\ref{4katt}, we use
$\blacktriangle$ and $\blacksquare$ to represent the states
responding to Exo-1 and Exo-2 respectively. Dotted lines are drawn
to show the dynamics and linkages of the two separated states. After
splitting, the nine states of the blue attractor combine to form the
four pairs of states of the red attractor in Fig.~\ref{4katt}(b),
the states in each pair being separated by a displacement of 1 in
both the horizontal and vertical directions. The states in each pair
move in the same direction simultaneously in the payoff space while
the signals in Exo-1 and Exo-2 are identical. Similarly, when an
error occurs in the red attractor, the four pairs of states may
split and combine to form the nine pairs of attractor states of the
green attractor in Fig.~\ref{4katt}(c), and the displacements in
these new pairs are 2 in both directions. The transition from the
green to pink attractor in Fig.~\ref{4katt}(d) can be described
similarly. In summary, the attractors are different in the vertical
and horizontal displacements separating the two groups of attractor
states. From Fig.~\ref{4katt}(b)-(d), the vertical and horizontal
separations between the two groups of attractor states are constant
for each attractor. Moreover, the separations change by 1 when an
attractor transition takes place due to the occurrence of an error.
Thus, the horizontal and vertical separations in the blue, red, green and pink attractors
are 0, 1, 2 and 3 respectively.

\begin{figure}
\centerline{\epsfig{figure=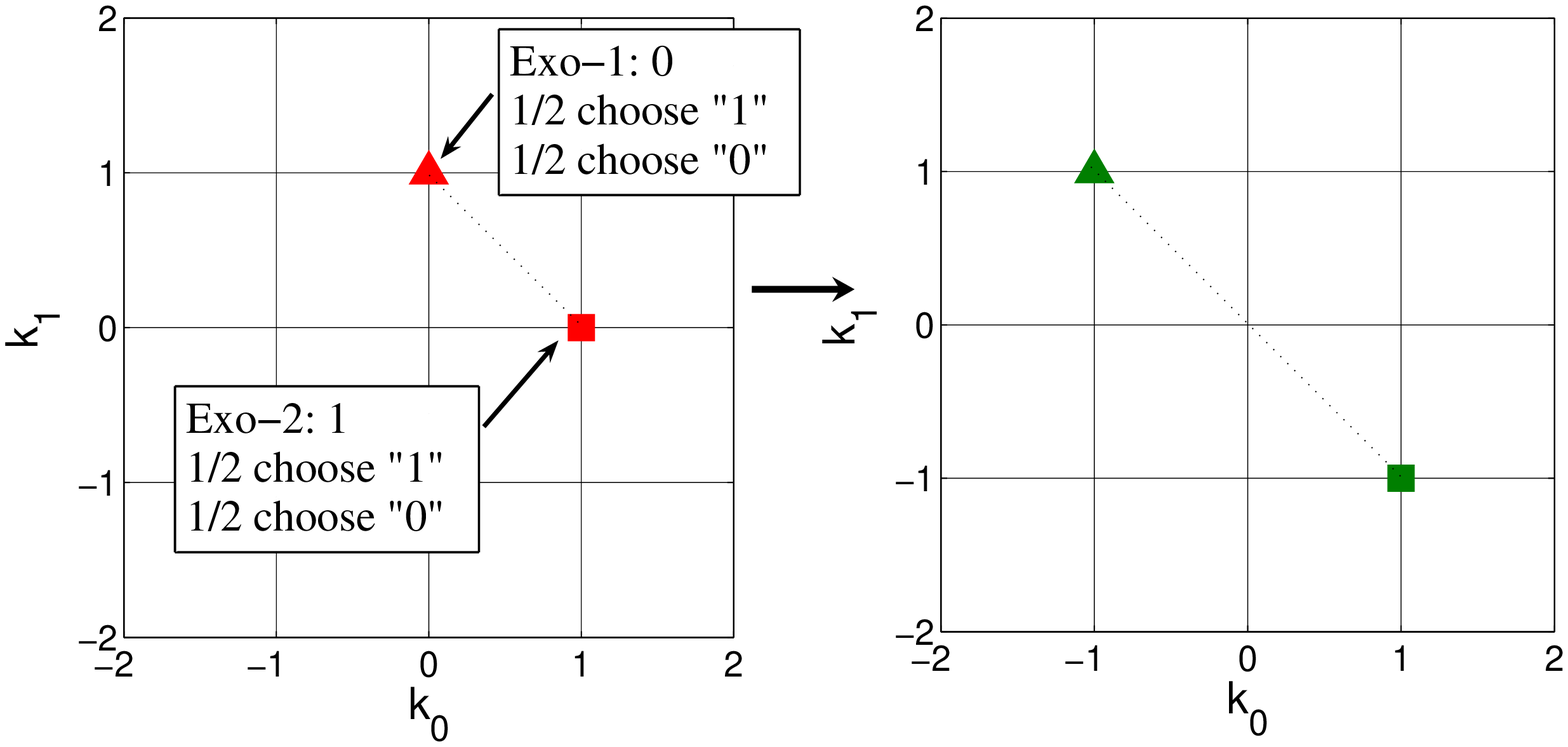,width=1.0\linewidth}}
\caption{\label{attTran} Mechanism of attractor transition. If the
minority side is 0, both attractor states decrease by 1 and transit
to another attractor as shown in the right hand side.}
\end{figure}

The mechanism of attractor transition is explained in
Fig.~\ref{attTran}. Suppose the attractor states responding to Exo-1
and Exo-2 lie on the $k_1$ and $k_0$ axes respectively, and the
corresponding input signals are 0 (for Exo-1) and 1 (for Exo-2). In
this case, $k_0=0$ among the agents responding to Exo-1. From
Eq.~(\ref{payoff}), all strategies have the same cumulative payoff.
Since the strategies are randomly picked, on average of half of the
agents in group choose 0 and the remaining half choose 1 to the
lower order. A similar argument applies to the agents responding to
Exo-2. The winning decision will then be determined by the
sample-dependent fluctuations of the strategies to the next order.
The minority side is 0 with probability 1/2 so that the locations of
the groups $\blacktriangle$ and $\blacksquare$ in Fig.~\ref{attTran}
both decrease by 1 along the $k_0$ and $k_1$ axes respectively.
Hence, the displacements between these two groups increase by 1,
transforming the attractor from red to green.

%%%%%%%%% Figure 5 %%%%%%%%%%%%
%%%%%%%%%%%%%%%%%%%%%%%%%%%%%%%%

%%%%%%%%% Figure 6 %%%%%%%%%%%%
\begin{figure}
\centerline{\epsfig{figure=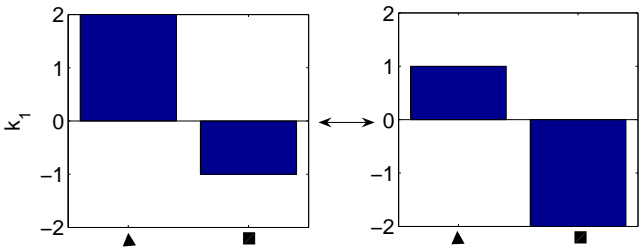,width=1.0\linewidth}}
\caption{\label{barchart} The alternation of $k_1$ between (2,-1)
and (1,-2) of a pair of states in the pink attractor. (Left) For signal 1, the $\blacksquare$ group has more agents favoring decision 0 since $k_1=-1$, but the $\blacktriangle$ group has even move agents forming 1 since $k_1=2>|-1|$. The minority side is 0. (Right) The minority side is 1.}
\end{figure}
%%%%%%%%%%%%%%%%%%%%%%%%%%%%%%%%

From the above results, we can deduce that the attractor transitions
do not have infinite stages. The final stage is the pink attractor
as shown in Fig.~\ref{4katt}(d). The vertical and horizontal
separations between a pair of states in this attractor are 3. Both
the $k_0$ and $k_1$ coordinates of the pair of states alternate
between (2, -1) and (1, -2) in the pink attractor, as shown in
Fig.~\ref{barchart} and as verified analytically. The physical
explanation for this stoppage of further attractor transitions is
contributed by the intrinsic {\it antipersistent} nature of Minority
Game. Antipersistence is the response of the system to a given
signal being opposite to that of the signal's previous
occurrence~\cite{Metzler}. This nature resists consecutive winnings
or losings for the same signal, causing the coordinates of the pink
attractor states to oscillate between 2 and 1, or -1 and -2, but
never reach 0 where transitions to further attractors are possible.
Therefore, it limits the number of attractors of this complex system
under noisy signals to 4. Thus, antipersistence strengthens the
stability of the whole system and avoid large fluctuations.

\section{\label{sec:level6} Conclusion}

In this paper, we have introduced a new variation of Minority Game,
``Minority Game with Errors''. The new feature is the usage of two
channels of signals to model the effects of errors occurred in the
input signals. We found that the noisy signals induce transitions
between different attractors, but the number of accessible
attractors is limited by antipersistent effects. Hence,
antipersistence is able to stabilize dynamical systems, making them
robust against noise.

Antipersistence is a generic feature which can be extended beyond a
single model. In financial markets, a lot of irrelevant information
or news may bother decisions of an individual. Nevertheless, the
whole system will not be affected by noisy signals easily. Similar
to the phenomena that we studied in this model, antipersistent
nature stabilizes the system by restricting the number of stages in
attractor transition. Therefore, the systems are prevented from
breaking down.

It would be interesting to explore further the manifestation of
antipersistence in the extensions of present model. For example, we
may consider the effects of more complicated errors using more
channels and strategies, the reception of noisy channels with
different probabilities for different agents, and the stabilization
effects of diversity in the initial references.

In summary, the Minority Game with multiple channels is a realistic
model for physicists to explore the importance of information or
errors to complex systems.

We thank S. W. Lim and C. H. Yeung for meaningful discussions. This
work is supported by the Research Grant Council of Hong Kong (DAG
05/06.SC36).

\end{document}